\documentclass[a4paper,doc,12pt, natbib, endfloat]{apa6}

\usepackage[english]{babel}
\usepackage[utf8x]{inputenc}
\usepackage{amsmath}
\usepackage{setspace}
\usepackage{pdflscape}
\usepackage{graphicx}
\usepackage[colorinlistoftodos]{todonotes}

\usepackage{geometry}
\usepackage{hyperref}
\usepackage[most]{tcolorbox}
\usepackage{times}

\usepackage{amsfonts}
\usepackage{amssymb}
\usepackage{listings}
\makeatletter
\def\Let@{\def\\{\notag\math@cr}}
\makeatother

\usepackage{amsmath,amssymb}

\usepackage{subcaption}
\usepackage{tabularx} 
\newcolumntype{L}[1]{>{\raggedright\arraybackslash}p{#1}}
\newcolumntype{C}[1]{>{\centering\arraybackslash}p{#1}}
\newcolumntype{R}[1]{>{\raggedleft\arraybackslash}p{#1}}
\usepackage{threeparttable} 
\usepackage{siunitx} 
\usepackage{pdflscape} 
\usepackage{hhline} 
\usepackage{tabularx} 
\usepackage{booktabs} 
\usepackage{color}
\usepackage{hyperref}

\usepackage[
singlelinecheck=false 
]{caption}

\title{
The validity of RFID badges measuring face-to-face interactions
} 
\shorttitle{RFID validity}
\author{Timon Elmer\footnote{Corresponding Author: Timon Elmer, \href{mailto:timon.elmer@gess.ethz.ch}{timon.elmer@gess.ethz.ch} \\ 
cite as: Elmer, T. Chaitanya, K., Purwar, P., \& Stadtfeld, C. (2019). The validity of RFID badges measuring face-to-face interactions. \textit{Behavior Research Methods.} Advance online publication. \href{https://doi.org/10.3758/s13428-018-1180-y}{https://doi.org/10.3758/s13428-018-1180-y}
}, Krishna Chaitanya, Prateek Purwar, \& Christoph Stadtfeld}
\affiliation{Chair of Social Networks, ETH Zürich, Switzerland
 }
\abstract{
Face-to-face interactions are important for a variety of individual behaviors and outcomes.
In recent years a number of human sensor technologies have been proposed to incorporate direct observations in behavioral studies of face-to-face interactions.
One of the most promising emerging technologies are active Radio Frequency Identification (RFID) badges. They are increasingly applied in behavioral studies because of their low costs, straightforward applicability, and moderate ethical concerns.
However, despite the attention that RFID badges have recently received, there is a lack of systematic tests on how \textit{valid} RFID badges are in measuring face-to-face interaction. 
With two studies we aim to fill this gap. Study~1 (N = 11) compares how data assessed with RFID badges correspond with video data of the same interactions (construct validity) and how this fit can be improved using straightforward data processing strategies. The analyses show that the RFID badges have a sensitivity of 50\% that can be enhanced to 65\% when flickering signals with gaps of less than 75 seconds are interpolated. The specificity is relatively less affected by this interpolation process  (before interpolation 97\%, after interpolation 94.7\%) -- resulting in an improved accuracy of the measurement. In Study~2 (N = 73) we show that self-report data of social interactions correspond highly with data gathered with the RFID badges (criterion validity).
 
}

\begin{document}

\maketitle

\newpage

\section{Introduction}

Face-to-face social interactions are a central activity in human lives and the desire to socialize with others is a core motivation for human behavior \citep{Baumeister1995}. Face-to-face interaction (or the lack thereof) have been linked to diverse outcomes such as psychological well-being, creativity, and success \citep[e.g.,][]{Steger2010, Lechler2001, Kawachi2001, Perry-Smith2006, Reis2000a}. 
In many contexts, it is thus important  to understand how often, under what circumstances, and with whom individuals engage in such social interactions.


However, face-to-face interactions have been difficult to measure. Self assessments of interactions suffer from known biases such as duration neglect  \citep{Fredrickson1993} or recency effects \citep{Greene1986}. 
Because of such shortcomings, \citet{Baumeister2007} have advocated for direct observations of behavior. 
Direct behavioral observation studies can indeed overcome problems of individual biases but are typically limited to small social contexts and a short observation periods.
Observational studies that make use of video recording and automated image recognition (possibly combined with automated speech recognition) hold great promise for scaling up observational studies \citep[e.g.,][]{Haritaoglu2000}. Automated recordings through camera and speech capture a lot more information than merely face-to-face interactions, for example, about expression of emotions, conversational content, or about individuals who are not participating in a study. Where such measures are not of key interest, this raises ethical concerns and methodological challenges. Privacy and informed consent (of incidentally recorded individuals) are hard to achieve. 
This is in line with the observation of \citet[][p. 399]{Baumeister2007} that ``sometimes, observations are unethical, unfeasible, or impossible''.
Smart-phone based measures are more suitable to only involve informed participants. They have been proposed to collect data of social interactions \citep{Miller2012}. Built-in sensors such as GPS \citep{Ashbrook2003}, WiFi \citep{Sapiezynski2017}, or Bluetooth \citep[e.g.,][]{Eagle2006} can help to identify the spatial co-location of individuals and electronic forms of interaction. However, the resolution of such technologies is too rough to allow identifying when people face each other in a social interaction, as they can capture at most who is in the same room.

\begin{figure}[ht]
    \includegraphics[scale=0.3]{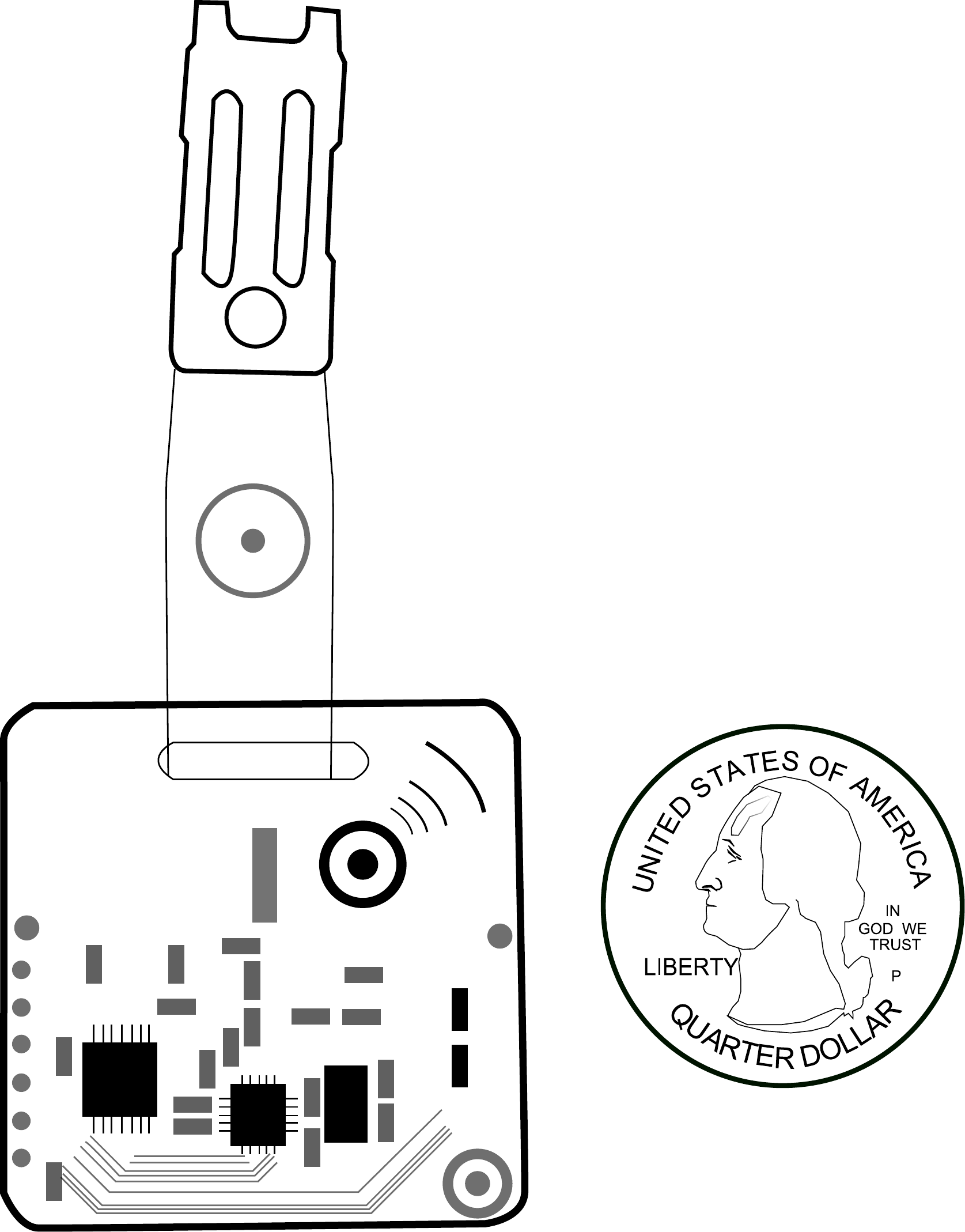}
    \caption[tag]
      {An illustration of an RFID badge and a quarter dollar for size comparison.}
    \label{RFIDpicture}
\end{figure}

\begin{figure}[ht]
    \includegraphics[scale=0.55]{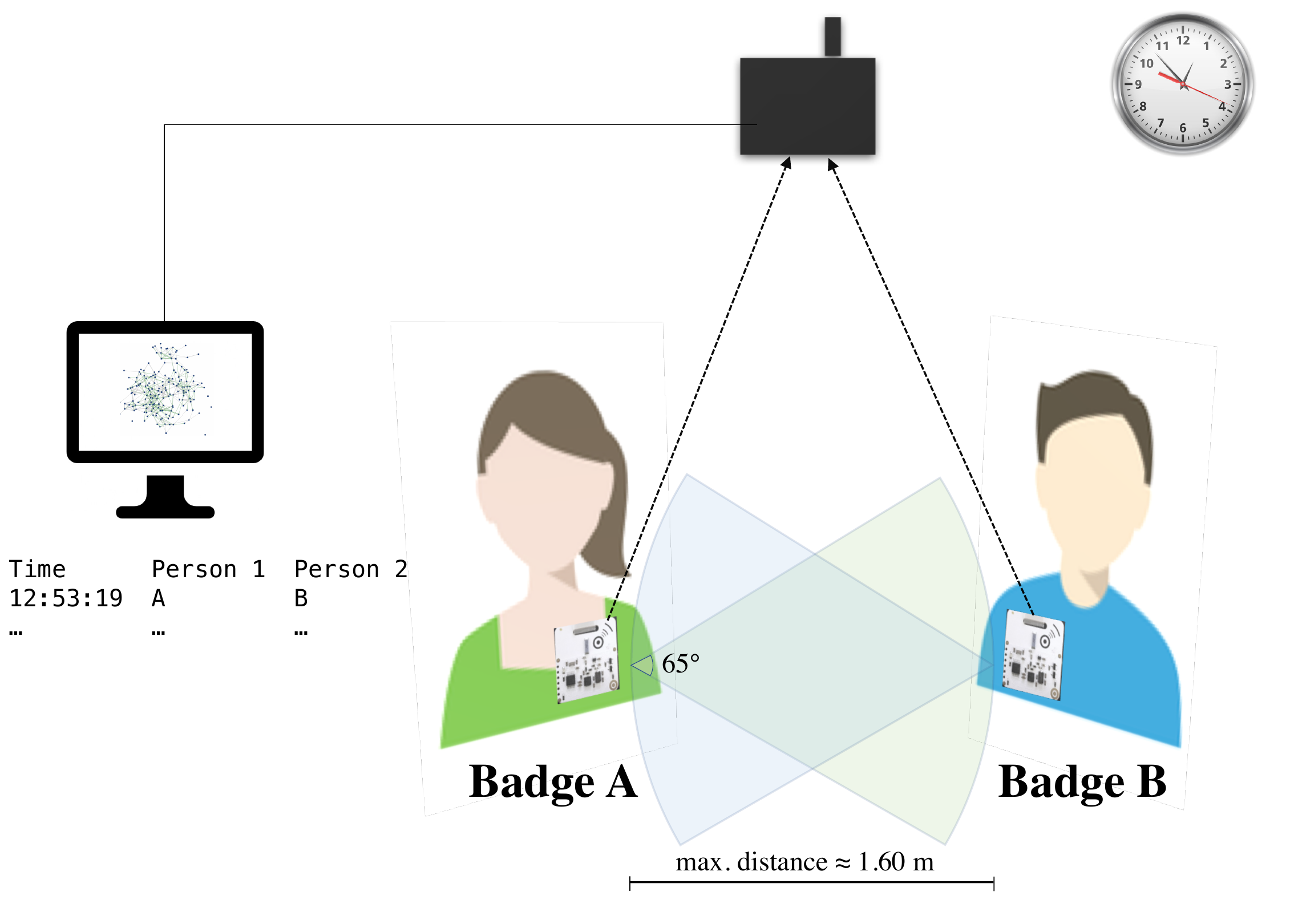}
    \caption[illustration]
      {An illustration of a face-to-face interaction measured with RFID badges. In this example badge B detects the signal from badge A and sends the information to the RFID reader that then passes that information on to the server computer. The time of the interaction together with the information that badges A and B are interacting is recorded. Details on the tests determining the distance and angle for signal detection between two badges can be found in Section~\ref{pretest}.}
    \label{illustration}
\end{figure}

One of the most promising proposals for the collection of face-to-face interaction are sociometric badges \citep{Pentland2008} that can be experimentally applied to collect data within bounded settings, such as within organizations, schools, or at conferences \citep[e.g.,][]{Waber2010,Pachucki2014,Scholz2013,ElmerDSSI}. These sensors are worn by study participants and automatically record when two study participants face each other in close physical distance. 
A technology that has been used in many applied studies are sociometric badges based on Radio Frequency Identification \citep[RFID;][]{Cattuto2010, Lederman2017}. Figure~\ref{RFIDpicture} shows a sketch of an RFID badge and its size.
RFID badges are typically worn on the chest by study participants (possibly hidden under a name tag) and measure if another study participant's badge is in short proximity (up to 1.6 meters) and in an angle that indicates that those two people are facing each other (each badge scans an angle of about 65 degrees). The measurements of two RFID badges are recorded in real-time by stationary routers that can capture broadcast signals from badges within a certain radius (depending on the architectural layout); the space in which interactions are recorded thus needs to be defined and tested in advance. The interaction data are then stored on a database server. Figure~\ref{illustration} illustrates the minimal setup of an RFID study. 
Section~\ref{sec:rfid-badges} and earlier literature \citep{Cattuto2010, Want2006} introduce additional technological details. 

RFID badges have been applied in diverse behavioral research studies, for example, to investigate social interaction patterns in a hospital \citep{Isella2011a}, at conferences \citep{Cattuto2010}, or to understand how social interactions are associated with well-being of a company's employees \citep{Chancellor2017} or of school children \citep{Pachucki2014}. The chief advantages of RFID badges as compared to more complex social sensor systems are that i) the collected data are minimal and do not capture video, speech or behavior of non-participants, reducing the risk of unnecessary privacy intrusion, ii) they are relatively cheap to assemble and thereby allow scaling up data collections to larger contexts and longer periods, iii) the technology is based on established industry standards which means that it can be considered more robust than purpose-built sensors, and iv) they can be employed in varying contexts in which face-to-face interactions are to be studied. The small scale of the badge permits that they can, for example, be integrated in a name tag. We think that this technology holds great promise in psychology and other fields of behavioral sciences, in particular due to recent advancements in statistical methodology for the study of time-stamped interaction data \citep{StadtfeldBlock2017, Stadtfeld2017, Stadtfeld2011a, Pilny2016, Butts2008}. 

A crucial question, however, remains to be answered: Do RFID badges actually measure what they are expected to? In stark contrast to the increasing use of RFID badges in research studies, there is a lack of thorough validation studies. Some functional tests have been proposed \citep[e.g.,][]{Cattuto2010, Isella2011a} that, however, mostly focus on technical functionality in lab settings. 
We replicate a number of those lab tests and provide guidelines that relate to the detection range, the detection angles, and the role of batteries on the technical validity in Section~\ref{pretest} (Pretests).
Measurement validity needs to be evaluated in the field as individual behavior has a direct effect on the measurement: The distance at which individuals communicate, their body angles, whether they are moving, how many people are interacting, hand gestures, or the presence of other objects (e.g., holding a glass of water) may all affect measurement quality. This paper closes this gap 
by proposing validity tests based on real-life data. 
It assesses two different types of validity. In the first study we assess the \textit{construct validity }of RFID badges by comparing social interaction data gathered through RFID badges with human-coded video data of the same interactions in a small setting (N = 11 individuals; 76.7 minutes of data recorded). We first assess the overall construct validity of the RFID data (i.e., the overlap between video and RFID data). We then test how the validity can be improved by imputing missing data that stem from signal instability and provide practical guidelines on data preprocessing strategies. In the second study we test the \textit{criterion  validity} of RFID badges by assessing how social interactions measured with RFID badges correspond with self-report measures of social interactions. The second study is situated in a larger context and involves N = 73 individuals and 36.86 hours of interaction data.

\section{Pretests}\label{pretest}

Before validating the RFID badges in field experiments (Study 1 and Study 2), we conducted a number of tests assessing the geographical ranges in which the RFID badges and the RFID readers operate. All tests were carried out with the same set of five badges. We sought to answer three questions: (1) What is the badge and edge detection-range of the readers (i.e., the distance from the reader within which the presence of a badge is detected and the distance within the presence of a signal between to badges is detected)? As there were no signal differences between readers, we report only results of the tests conducted with one reader. When there was no object in between the badge and the reader (e.g., a wall or a person shielding the signal), badges were detected more than 50 meters away from the reader. The presence of a person between the badge and the reader reduced the detection-range to 26.8 meters (SD = 0.54). Walls that are in between the reader and badges reduce the range depending on wall thickness and construction material. Hence, we advise to test reception ranges within the spatial setting in which interactions should be recorded before the collecting data. (2) What is the edge detection-range between two RFID badges? To answer this question, we took five random pairs of badges and tested up to which distance and angle two badges would detect each other (i.e., measure a social interaction). On average, at up to 1.61 meters distance (SD = 0.35) edges between two badges were detected. Depending on the material that is placed on top of the RFID badge (e.g., a plastic name tag holder) this range is reduced. A single layer of paper (that can be used as a name tag) had no significant effect on the detection range. The angle between two RFID badges for an edge to be detected on average was 32.6 degrees (SD = 7.56) from the horizontal and vertical zero-axis (in total about 65 degrees towards all sides). We found no effect of the distance between the RFID badges and the reader on the edge detection-range. (3) Do battery properties affect the edge detection-range of two RFID badges? There is no effect of battery run-time on the edge detection-range (tested up to 72 hours). We have encountered lower detection-ranges of batteries (of type Panasonic coin lithium batteries CR2032) that had been used two months in advance and were stored properly in between. Such differences were not found for batteries that had only been used for a week. Hence, we recommend using temporally new batteries when collecting data.

\section{Study 1}

Study 1 aims at testing the construct validity of RFID badges by assessing how interactions measured with the RFID badges correspond to human-coded interactions of video data.
Moreover, we evaluate if straightforward data processing strategies can enhance the validity of the RFID badges. Those strategies take into account that missing data are often systematic and, for example, are characterized by fluctuating stability of the signal within a dyadic interaction. In particular, we test three such strategies that relate to i) the duration of social interactions, ii) the time between two interactions and iii) triadic configurations. 

In some earlier studies, scholars have restricted themselves to analyzing time windows of a given length \citep[e.g., 20 seconds; ][]{Cattuto2010} without considering how long the signal was recorded for within that time window.
The reasoning behind that threshold is that a sensor may pick up the signals of other sensors in situations that are not a face-to-face interactions, for example, when two interacting groups of individuals are standing in close proximity or when individuals pass each other while moving through the crowd. 
In the past, scholars have investigated how various cutoff-points of social interaction duration can be used to predict future interactions \citep[e.g.,][]{Scholz2012} or self-reports of social interactions \citep[e.g.,][]{Atzmueller2018,  Smieszek2016}. To the best of our knowledge, no study yet has investigated how these cutoff-points affect the construct validity of the RFID badges to measure face-to-face interactions. 
Hence, in a first step, we test how variations of this threshold (rather than taking the ad-hoc threshold of 20 seconds) contribute to the validity. 

\begin{figure}[ht]
    \includegraphics[scale=0.8]{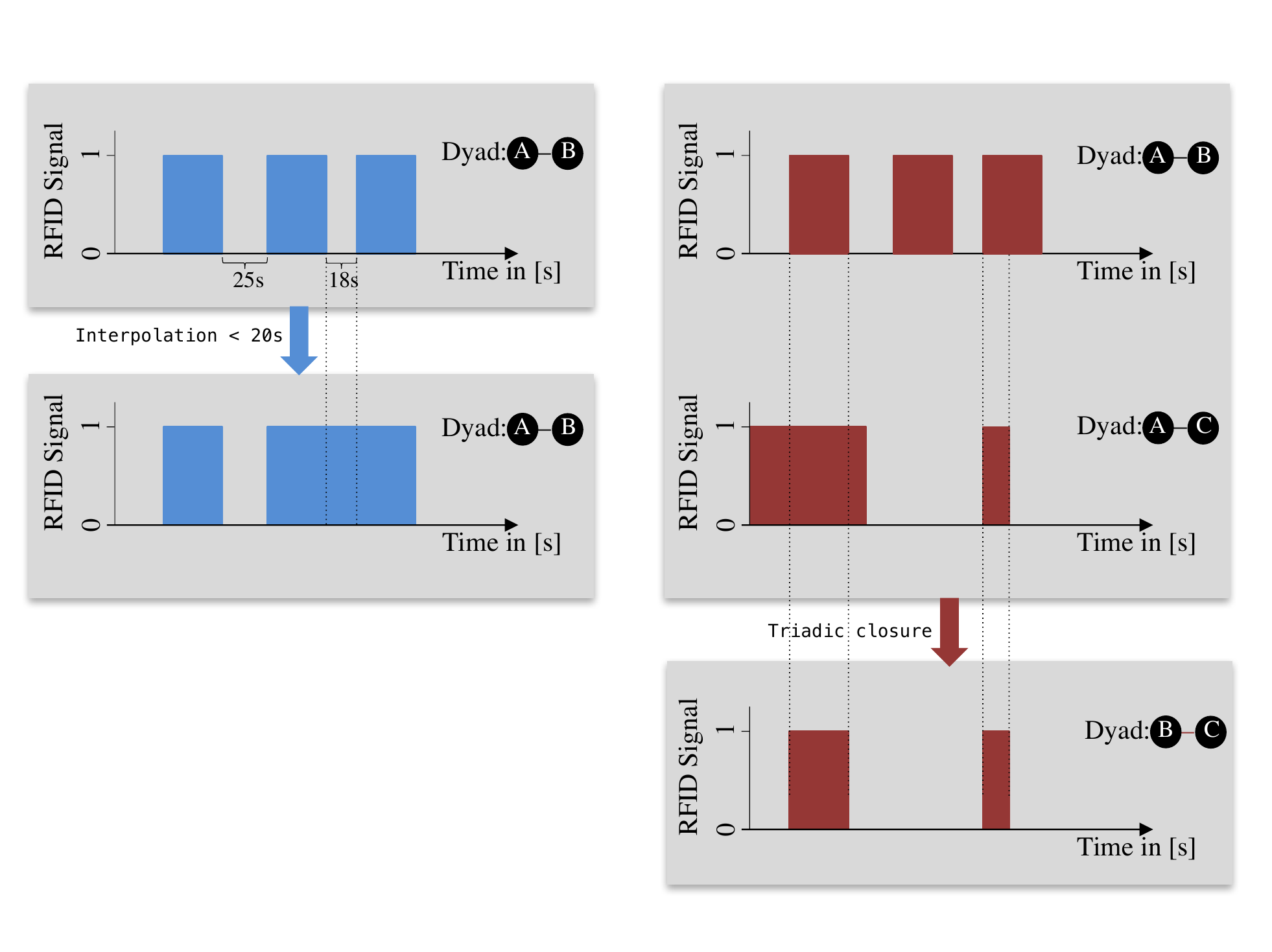}
    \caption[functions]
      {Left: An illustration of the \textit{interpolation }data processing strategy, in which interactions of the same dyad are interpolated if they are maximally $x$ seconds apart (here $x = 20$). Right: An illustration of the \textit{triadic closure} data processing strategy, in which a tie between B and C is added.}
    \label{fig:functions}
\end{figure}

Second, we assess to what extent merging signals between two individuals, with respect to how long these interactions are apart, improve the validity. The reasoning behind this merging strategy is that even if individuals are involved in a longer face-to-face interaction, their body movements, or interfering objects such as other individuals passing by or drinking glasses may interrupt the signal at times. With this strategy, we make use of the continuous and fine-grained data to overcome such measurement biases. Merging two signals into one measure may thus increase the validity of the measure. 
We refer to this strategy as \textit{interpolation}. Figure~\ref{fig:functions} (left panel) illustrates this data processing strategy. 

Third, we test the contribution of adding missing ties in interaction triads. For instance, if individual A is interacting with B and C at the same time, then a tie between B and C is added for the time in which A is in interaction with B and C. Because social interactions are \emph{unimodal} (i.e., individuals can only engage in one interaction at a time), we may assume that B and C also interacted with each other if A is interacting with both of them, even if there is no, or no stable signal between them. 
Due to the nature of the RFID technology, interactions are only observed if the individuals are facing each other. Hence, if two people (B and C) are standing next to each other because they are listening to A, a tie between B and C would not be detected. Furthermore, in larger groups, the narrow angle of the RFID badges might not allow the capturing of each pair of individuals involved.
Closing triads might therefore improve the validity of the RFID badges.  Figure~\ref{fig:functions} (right panel) shows the \textit{triadic closure} data processing strategy. 

 \subsection{Methods Study 1}
 
\subsubsection{Design} \label{sec:design}

\begin{figure}[ht]
    \includegraphics[scale=0.8]{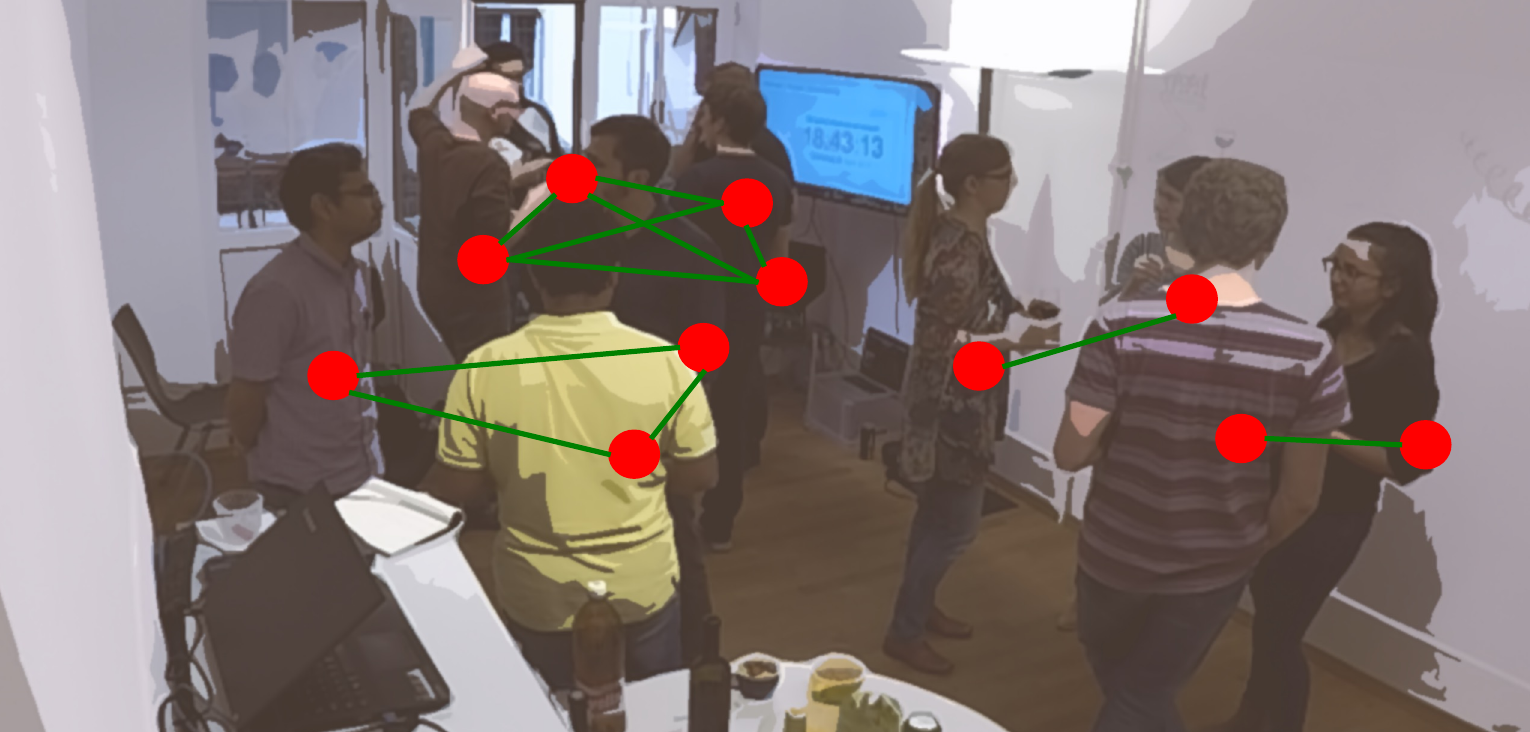}
    \caption[screenshot]
      {A screenshot (with an anonomyzing blur) from the video recording including graphical representations of social interactions.}
    \label{screenshop}
\end{figure}

Staff and students of a Swiss university were invited to the experiment, advertised as an after-work get-together event. 11 individuals took part in the experiment. 
Three participants were female. A room of about $20m^2$ was set up with a camera (model: GoPro 4) covering the whole room and two RFID readers (devices to detect the signals between RFID badges in real time) situated in two opposite corners of the room. The density (individuals per square meter) was chosen to be similar to Study~2 (Section \ref{sec:study2}). 
Upon arrival, each participant was equipped with an RFID badge and was instructed to wear it on the top layer of clothing at chest height. No further instructions were given to the participants. As expected, participants engaged in social interactions with other participants. During the event beverages and snacks were served. Figure~\ref{screenshop} illustrates the setup of the experiment. 60\% of the pairs of participants knew each other beforehand. A total of 76.7 minutes of video and RFID-data were recorded. Summed  over all pairs of individuals ($N_{\text{pairs}} =  \frac{N(N-1)}{2} =55$) , there are 55 * 76.7 minutes ( = 70.3 hours) of dyadic data recorded. To compare the video data to the RFID data, we then transformed this dyadic data structure to a linear time dimension for each pair of badges, indicating if for a given second an interaction was present or not (0 = no, 1 = yes). Further details on the comparison are given in Section \ref{sec:fit}. 

\subsubsection{RFID badges}
\label{sec:rfid-badges}

RFID badges were used to capture social interactions between the participants of the experiment.
The firm- and software for the RFID badges and readers that are used in this article were taken from the OpenBeacon project (www.openbeacon.org). OpenBeacon is an open source software and hardware project. Similar technologies have been proposed in other projects and we expect those to behave similarly \citep{Lederman2017, Cattuto2010}\footnote{In addition, the SocioPatterns project (www.sociopatterns.org) has tweaked the hard- and software to improve the performance of the badges in particular settings.}. Our version of the OpenBeacon software
can be downloaded from \href{www.osf.io/rrhxe}{\path{osf.io/rrhxe}}.

A set of 11 active 2.4 GHz RFID badges (nRF24L01P chipsets) that  uses the proprietary Nordic Semiconductor radio protocol were used in this experiment.
The OpenBeacon proximity firmware was used to track the location of the badge as well as interactions once a contact between two badges is established via the regularly transmitted beacon packets that the badge constantly sends. The packets containing the information about position and proximity of the badge are received by the nearest OpenBeacon Easyreader PoE II (for brevity called reader) which sends the information to the server infrastructure via LAN cable.  
Our experimental setup consisted of 2 readers, 11 RFID badges for the 11 participants and a computer which acts as a server. The readers were directly connected to the server computer, which collects and stores information received by the RFID readers on the level of a fifth of a second.
The reader scans its environment five times every second.  The server then receives information about which badge is detected by which reader and between which badges interactions are recorded. Badges can record multiple interactions at the same time. We then transformed these data into a time-stamped edgelist, of which the first 6 interactions are shown in Table~\ref{data}. For instance, the first row of the table indicates that the badges with IDs 3 and 5 interacted with one another from 18:19:46 to 18:19:58.

The firm- and software as well as the schematic and hardware design of the RFID badges are freely available on the website of the OpenBeacon project (www.openbeacon.org). Using this open source information, the badges can be assembled by interested research groups. Other developers of similar hard- and software also provide their source code online \citep[e.g., ][]{Lederman2017}. More details on the RFID technology and its application to measure social interactions can be found elsewhere \citep{Cattuto2010, Want2006}. 

\begin{table}[ht]
 \caption{Example data collected with RFID badges}
  \label{data}
\begin{tabular}{llll}
 \hhline{====} 
Start & ID Badge A & ID Badge B & End \\
\hline
18:19:46 & 3 & 5 & 18:19:58 \\
18:19:47 & 1 & 10 & 18:20:15 \\
18:19:47 & 1 & 8 & 18:22:32 \\
18:19:49 & 10 & 8 & 18:22:35 \\
18:19:53 & 2 & 5 & 18:20:37 \\
18:20:04 & 6 & 11 & 18:20:14 \\
\hhline{====} 
\end{tabular}
\end{table}

\subsubsection{Human-coded interactions}

Goffman (1956, p. 18)\nocite{Goffman1956} defined a face-to-face social interaction as the “reciprocal influence of individuals upon on another's actions when in one another's immediate physical presence”. For our setting, this definition is too broad as it, for example, may include physical presence in the same room. Hence, we narrowed this definition and coded a face-to-face interaction when two individuals were talking or listening to each other or when they were part of the same group conversation. More specifically, an interaction was coded when an individual directed his/her attention as indicated by the body movement (turning of head and/or rotating the body) to another person or group for more than 10 seconds. Briefly turning one's attention (< 10s) to someone else or another (interaction)-group was not coded as an interaction. Similarly, leaving the interaction for less than 10 seconds (e.g., to put down a drinking glass to the nearest table), was not considered as two separate interactions. In group conversations, every group member was coded as interacting with every other member, irrespective of the role in the group (i.e., speaker or listener). Interactions were coded to match the format of the RFID data, as shown in Table~\ref{data}. 

Two confederates independently coded the interactions in the videos. An overlap of 13.5 minutes (18\% of the total duration) was coded by both raters to compute the interrater reliability of the video coding. To evaluate the interrater reliability we computed Cohen's $\kappa$ \citep{Cohen1960}, which has been proposed as a chance-corrected agreement between two raters \citep{Hallgren2012}. 
Cohen's $\kappa$ was .96, indicating a very high interrater reliability \citep{Landis1977}. We therefore can assume that the two raters were consistent in their understanding of what constitutes a social interaction. The interactions of both raters were merged so that both raters accounted for half of the time coded.

\subsubsection{Assessing the fit between the video and RFID data}
\label{sec:fit}
When assessing the validity of a measure, the \textit{sensitivity} and the \textit{specificity} are the most prominent indices. We use the human-coded video data as the ground truth. 
Hence, in our case, the \textit{sensitivity }is defined as the true positive rate. In other words, sensitivity is the proportion of human rated interactions that are correctly identified as such by the RFID badges. \textit{Specificity} (the true negative rate) is the proportion of human-coded non-existent interaction that are correctly identified by the RFID badges. 
Formally, the sensitivity and the specificity are defined as $\frac{\text{TP}}{\text{TP} + \text{TN}}$ and $\frac{\text{TN}}{\text{TN} + \text{FN}}$, respectively. The classification indices (true positives (TP), false negatives (FN), etc.) for our analysis are defined in equations~\ref{equ:tp} to~\ref{equ:tn}:

\begin{equation}
\text{TP} = \sum_{d = 1}^{D} \sum_{i = 1}^{S} I\{R^d_i + V^d_i = 2\} \\
\label{equ:tp}
\end{equation}
\begin{equation}
\text{FP} = \sum_{d = 1}^{D} \sum_{i = 1}^{S} I\{R^d_i - V^d_i = 1\} \\
\end{equation}
\begin{equation}
\text{FN} = \sum_{d = 1}^{D} \sum_{i = 1}^{S} I\{R^d_i - V^d_i = -1\} \\
\end{equation}
\begin{equation}
\text{TN} = \sum_{d = 1}^{D} \sum_{i = 1}^{S} I\{R^d_i + V^d_i = 0\} \\
\label{equ:tn}
\end{equation}

Vector $R^d$ (RFID) is a dummy vector with the length of the total observation period in seconds $S$ indicating whether an interaction of dyad $d$ among all dyads $D = \frac{N(N-1)}{2}$ was recorded with the RFID badges at the respective second. Vector $V^d$ (Video) is a dummy vector of the same length indicating whether an interaction of dyad $d$ was coded in the video at the respective second. The elements $i$ of the vector indicate an entry that relates to a specific second, $S$ is the last recorded second and thus the vector length. 
$I\{A\}$ denotes an indicator function for condition $A$ and returns one if the condition is true and zero otherwise.

For the process of finding optimal values for \textit{minimal duration}, \textit{interpolation} and the number of iterations for which \textit{triadic closure} is performed, we use a single index that entails a combination of all classification indices called the \textit{accuracy}. Accuracy ($a$) assesses the percentage of correctly identified instances (i.e., seconds) and is defined as:
\begin{equation}
a = \frac{\text{TP} + \text{TN}}{\text{TP} + \text{TN} + \text{FN} + \text{FP}}
\end{equation}
We choose to optimize this single index because one value can be optimized more easily and thereby weights every second equally compared to relative indices (such as sensitivity and specificity). Nevertheless, we also report the sensitivity and the specificity for a more detailed understanding of the validity. Alternative indices such as the sum of the sensitivity and specificity \citep{Koepsell1985} do not consider each correctly/incorrectly specified second equally, but relatively to the size of other cells in the classification table. Hence, we do not focus on such relative measures. Nevertheless, we provide robustness analyses for these indices in Appendix~\ref{app:S1_robust}.

\subsection{Results Study 1}

\subsubsection{Description of the data}

\begin{figure}[ht]
    \includegraphics[
    width = \textwidth 
    ]{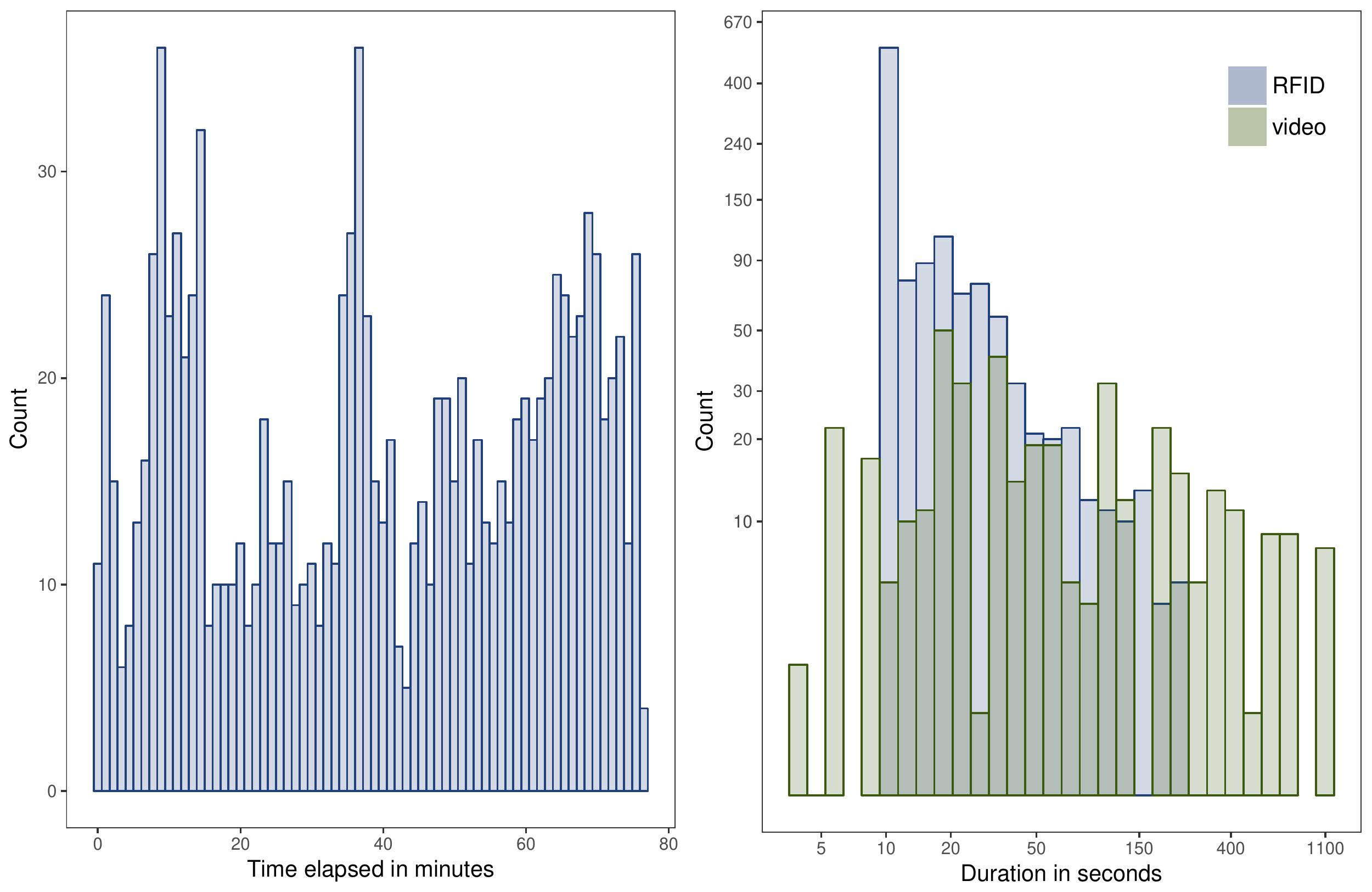}
    \caption[descriptives 1]
      {Left: Histogram of the number of RFID interactions for each minute elapse. Right: Histogram of the durations of the RFID and the video data with both axes log-transformed.}
    \label{hists}
\end{figure}

A total of 1,168 interactions with varying lengths were measured by the RFID badges. Figure~\ref{hists} (left) shows the overall RFID signal over time. It can be seen that the number of interactions changes through time with a maximum of 35 interaction pairs recorded. The maximum number of 55 interactions ($\frac{N(N-1)}{2}$) could have only been reached if all 11 participants had simultaneously interacted in one large group which never occurred. 
Figure~\ref{hists} (right) shows the durations of each interaction measured by the RFID badges and the video-coding. The interactions captured with RFID badges tend to be much shorter than the the video-coded interactions -- this could be an indicator that the signals relating to one interaction tend to be unstable\footnote{There are many interactions of length 10 seconds captured by the RFID badges, because we have set the "PROXAGGREGATION\_ TIME\_ SLOTS" parameter to 10 in the firmware of the badges. In pretests, lower values of this parameter caused some ''flickering'' of the signal (i.e., reoccurring edges within a small time window). Hence, interactions that are shorter than 10 seconds are stored as interactions of 10 seconds.}. For instance, while the video might record 5 minutes of an interaction between person A and person B, the RFID badge might record 5 unique 50-second-long signals of the same interaction between A and B. Because people tend to move their upper body during a conversation, the RFID signal might be interrupted from time to time (in this example: for 10 seconds every minute). We hope to reduce the flickering of the RFID signal with the interpolation processing or the transitive closure processing. When comparing the total duration of interactions (i.e., the sum of all interaction durations), the video-coded data records more interaction time (14.3 hours, 20.3\% of the possible 70.3 hours of dyadic data) than the RFID badges (8.4 hours, 12.0\%).

%

\subsubsection{Initial validity}

In this step we assess the fit between the interactions recorded through the RFID devices and the video-coded interactions. Table~\ref{initialstats} shows the classification table for this comparison. The classification table shows the number of seconds that were identified by the two methods (RFID and video) as positive or negative. The number of seconds that each of the two methods classified as a social interaction is denoted as positive in the classification table. Because the data is on a dyadic level, the number of seconds that need to be classified (as either positive or negative by the two methods) is the number of seconds that all dyads could have possibly interacted with one another (i.e., 55 dyads * 76.7 minutes, see Section~\ref{sec:design} for details). For instance, if the RFID badge and the video both indicate at a specific second that person A and person B interacted, then this adds one to the count of the top left cell of the table (true positives). But if the RFID badge does not indicate an interaction for that given second and the video data does, a count of one is added to the top right cell of the table (false negatives).

Based on the classification results of Table~\ref{initialstats} the sensitivity is 49.7\%, the specificity at 97.0\%, and the accuracy at 87.5\%, indicating that around 50\% of the seconds of interactions were detected by the RFID badges and 97\% of the seconds of non-interactions were correctly identified as such. This is a first promising observation because even if signals are unstable for a number of technical and behavior-related reasons, a signal that is captured by the RFID badges tends to be a reliable measure of an ongoing interaction and incidental measurements (false positives, the lower right cell of Table~\ref{initialstats}) seem to be rare.

\begin{table}[ht]
  \caption{Classification table (in seconds) of the initial comparison between the RFID data and the video-coded data}
  \label{initialstats}
  \begin{minipage}{\textwidth}
\begin{tabular}{
L{2cm}
L{2cm}
L{2cm}
L{2cm}
}
\hhline{====} 
\\
& & \multicolumn{2}{c}{RFID} \\
& & positive & negative \\
 \hline \\
Video & positive & 25,326 & 25,674 \\
& 	negative & 6,086 & 196,025 \\
\hline

\hhline{====}
\end{tabular}
 \end{minipage}
\end{table}

%

\subsubsection{Processing the data to improve the validity}

\paragraph{The contribution of a single data processing strategy}

\begin{figure}[ht]
    \includegraphics[scale=0.63]{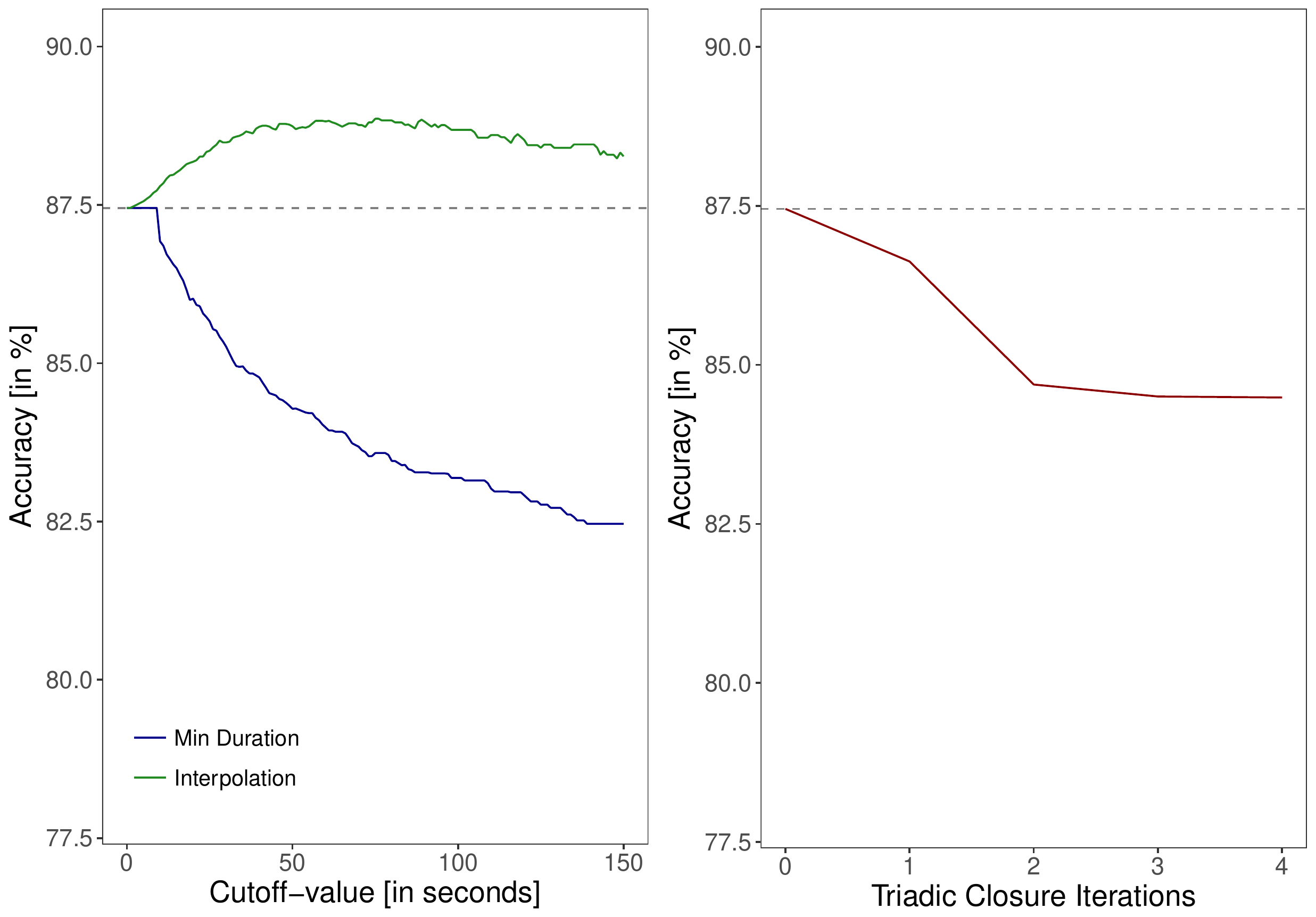}
    \caption[fit1]
      {Accuracy of the RFID data compared to the video coded data after the two strategies minimal duration and interpolation have been applied with different cutoff-values (left plot) and after several rounds of triadic closure iterations (right plot). The horizontal gray line indicates the accuracy of the unprocessed data.}
    \label{fit_1}
\end{figure}

Here we test the contribution of each of the three data processing strategies to the fit. Figure~\ref{fit_1} (left) shows the effect that deleting interactions which are shorter than a variable cutoff-value has on the accuracy. Each cutoff-value has a negative effect on the accuracy which indicates that deleting very short interactions (''flickering'') from the data is not a good strategy. Alternatively, one can merge two signals into one when there is only a short interruption between them (interpolation strategy).
The effect of this strategy is also shown in Figure~\ref{fit_1} (left). Merging interactions that are maximally 75 seconds apart showed the best fit, with an accuracy of 88.9\% (shown), and a sensitivity and specificity of 65.6\% and 94.7\%, respectively (not shown). Please note, that a whole range of values (50-100 seconds) are about equally good and clearly better than no interpolation. The detailed classification table of the interpolation strategy with a cutoff of 75 seconds is reported in the Appendix Table \ref{stats75}. 


Figure~\ref{fit_1} (right) shows the contribution of various iterations of triadic closure on the accuracy. 
Triadic closure is performed in iterations, as the imputation of one open triad may create new open triads that can be closed in a second interaction step. 
One iteration of triadic closure decreased the accuracy by 0.83\% (to 86.6\%) and resulted in a sensitivity of 61.4\% and a specificity of 93.0\%. Two iterations produced an even worse fit with an accuracy of 84.7\%, a sensitivity of 62.8\%, and a specificity of 90.2\%. A third and fourth iteration of triadic closure did not result in much further change of any of the fit criteria, which might indicate that by then no more triads can be closed and the network consists of fully connected interaction cliques. 

\paragraph{Combining data processing strategies}

In the next step we test how the sequential application of these strategies contribute to the fit. We do this by processing the data with the optimal value for the respective strategy and then applying the other two strategies to these data\footnote{We also used optimization algorithms to find optimal values for the three strategies jointly. Due to the vast number of local maxima, we were not able to find a optimal value with this method.}. So far, only the interpolation of interactions had a positive effect on the fit. Hence, there are no optimal values for the minimal duration and triadic closure strategy, for which we will instead use the cut-off defined by \cite{Cattuto2010} of 20 seconds minimal duration and one iteration of triadic closure. 

\begin{figure}[ht]
    \includegraphics[scale=0.63]{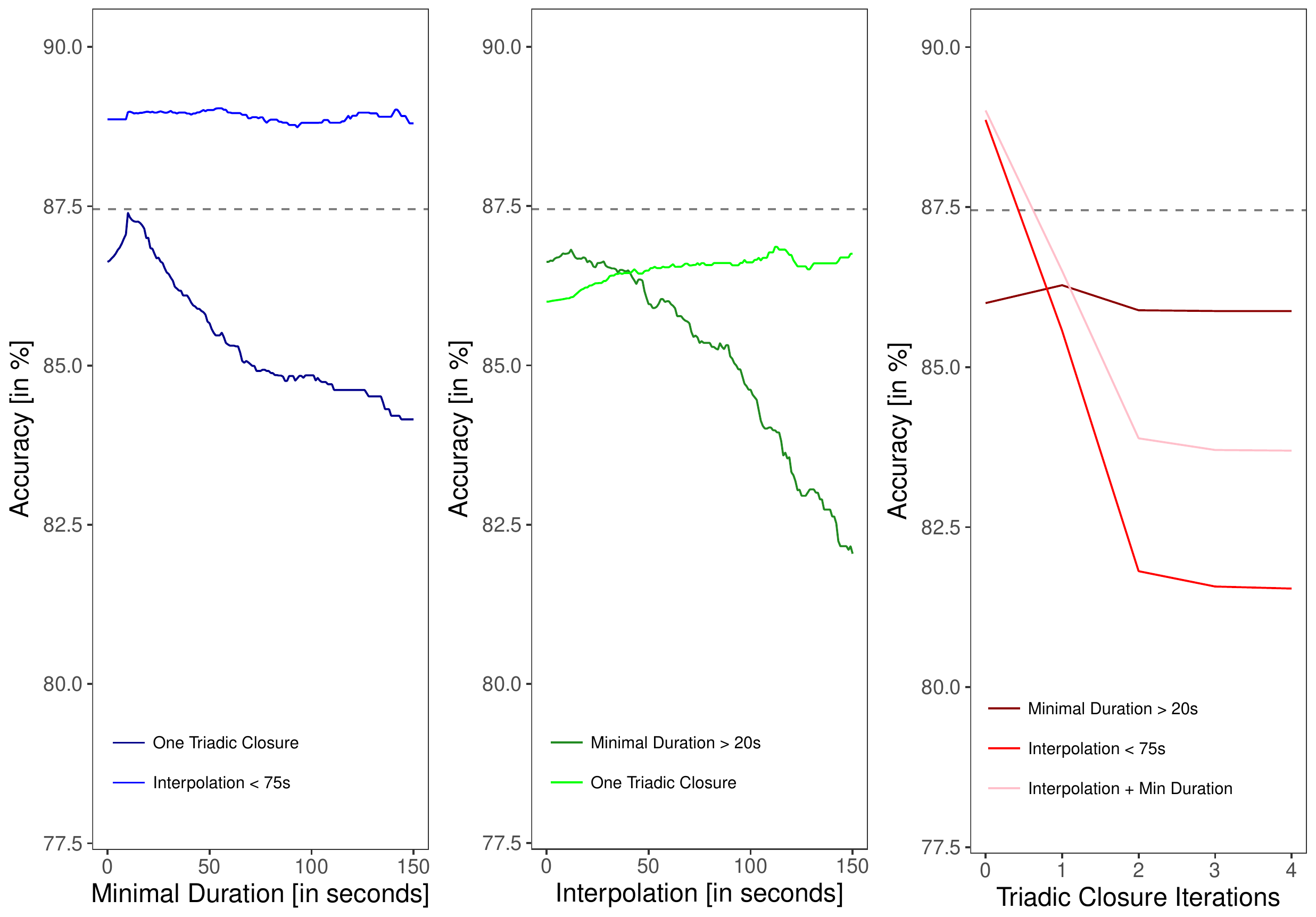} 
    \caption[fit2]
      {Accuracy of the RFID data compared to the video coded data with different cutoffs of the minimal duration (left), interpolation (middle), and different iteration rounds of triadic closure (right) applied to already preprocessed data. The horizontal gray line indicates the accuracy of the unprocessed data.}
    \label{fit_2}
\end{figure}

Figure~\ref{fit_2} shows the accuracy of the RFID data with two strategies sequentially combined. For instance, Figure~\ref{fit_2} (left) shows the effect on the accuracy of minimal duration after interpolation with a cut-off of 75 seconds or one triadic closure have been applied. The only combination of strategies that produced a slight (and potentially negligible) increase in the fit, was the deletion of interactions of 55 seconds duration after an interpolation of 75 seconds had been applied (see Figure~\ref{fit_2} left). 
We furthermore tested a combination of all three strategies with this new optimum. In other words, we closed triads on the preprocessed dataset with an interpolation of 75 seconds and duration deletion of 50 seconds. The results of this analysis (as shown in Figure~\ref{fit_2} right) reveal no improvement of the fit.

\subsection{Discussion Study 1}

In this study, we tested the construct validity of RFID badges to measure face-to-face social interactions by comparing data gathered with RFID badges to human-coded video data of the same social event. We have shown that the construct validity is reasonable, but not very high: 87.5\% of all seconds were identified correctly but about half the actual interaction-seconds were not recorded by the RFID badges (i.e., a sensitivity of 49.7\%). 
The specificity in particular tends to be very high (about 96\%) which indicates that if a signal between two badges is detected it is very likely that those two individuals actually interacted. 
The measure is thus rather conservative. 

We further tested how the application of three straightforward data processing strategies (minimal duration, interpolation, and triadic closure) contribute to the validity. We show that the validity of the RFID badges can be improved, with an accuracy of up to 89.0\% and a sensitivity up to 65.6\% when the interpolation criterion is used.  Thereby, a number of ``flickering'' interaction signals can be stabilized. This has the positive side effect that the number of recorded interaction events drops which may facilitate further statistical analyses.

Depending on the type of data, we recommend applying different data processing functions: If the network is rather sparse and researchers want to make sure to capture most of the interactions that actually happened (i.e., getting a high sensitivity, at the cost of the specificity), the data suggests to interpolate interactions with 50-100 seconds (with an optimum at 75 seconds). If it is important to capture the non-interactions precisely as well, then our advice is to interpolate with 75 seconds and then delete interactions shorter than 55 seconds for a slightly better fit than the formerly mentioned strategy.
Omitting interactions that are shorter than 20 seconds \citep[e.g.,][]{Cattuto2010} has a negative effect on the validity of the RFID data. This is not surprising, considering that most of the interactions were shorter than 20 seconds. Surprisingly, however, the strategy to close triads did not increase the validity. This indicates that measurements tend not to be much worse in the case of group interactions. 
Of the three strategies, only interpolating produced an improvement of the validity.


\subsubsection{Limitations}

Study 1 has a number of limitations. First, we validated the RFID badges in a very specific context (i.e., an after-work get-together event). The findings of this study might not carry across to other situations, where the room properties, culturally accepted personal distance \citep{Lomranz1976}, or previous relations between the participants are different. 
The physical distances (and therefore the RFID signals) between two individuals could be biased by, for instance, people's personality, cultural background, or substance consumption. 
However, we have specifically chosen this context because most applications of RFID badges (including our own study~2) have been conducted in very similar contexts where people that partially know each other engage in upstanding face-to-face interactions -- for instance, in a conference setting \citep{Cattuto2010}, museum exhibition \citep{Isella2011}, or a student welcome event \citep{Atzmueller2014}. 
Second, the majority of interactions that were measured by the RFID badges were shorter than 10 seconds but were rounded up to a duration of 10 seconds by the software. Distinguishing between very short interactions (i.e., less than 10 seconds) is therefore not feasible but also not within the scope of such a data collection, as RFID badges are intended to be applied to capture interactions in settings where participants spend several hours together. 

\section{Study 2}\label{sec:study2}

In the second study we aim to validate the measure of the RFID badges with self-report data of social interactions in a larger social context.  Moreover, we will assess how applying the three data processing strategies tested in Study 1 affect the fit between the RFID data and the self-report ratings. 

For this analysis, we collected social interaction data with RFID badges over the course of a welcome-weekend of students. After the weekend, we asked these students with whom they had pleasant social interactions. Hence, we will compare the RFID data with the self-reports of social interactions. 

\subsection{Methods Study 2}

\subsubsection{Participants}

The sample for Study 2 consisted of $N$ = 73 students. All of the weekend attendees agreed to wear the RFID badge during their waking times. 27 (37 \%) of the participants were female. These students just started studying at a Swiss university and were invited to spend a weekend together in a camp house in the Swiss mountains to get to know each other \citep[for details see ][]{ElmerDSSI, Stadtfeld2019}. Throughout the weekend, students participated in social activities that were intended to facilitate social integration. During the course of the weekend social interactions were assessed using active Radio Frequency Identification (RFID) badges. In the two days following the weekend, students were asked to participate in a survey that assessed with whom they had had social interactions. 51 (70 \%) of the participants administered this follow-up questionnaire. The institutional ethics board reviewed and approved this study. 

\subsubsection{Procedure}
Before the arrival on the remotely located camp house, each participant was equipped with a badge that consisted of the RFID badge and a piece of paper with their name printed on it. Participants were briefed on the badge's functionality and purpose of application. All participants were instructed to wear the RFID badge during their waking times and place them at chest height. During the event, this was checked by study confederates, who instructed the participants to wear the badge correctly. 
Throughout the data collection (Friday 7pm to Sunday 8am), some group activities  took place (e.g., group games, lectures), but most of the time was unstructured so that participants could interact with each other or play games. At night parties were organized by a student organization, during which many participants consumed alcoholic beverages. Data recorded during collective events such as talks by university professors, or in time windows when most participants were asleep were not treated as interaction data in this study.

\subsubsection{Materials}

\paragraph{RFID badges}

The same RFID badges as used in Study 1 were distributed among the participants (i.e., active 2.4 GHz Radio Frequency Identification devices). Before arrival of the participants, the three leveled camp house was equipped with 8 RFID readers so that in every room of the house and in commonly used outside areas (e.g., smoking area) signals between RFID badges could be detected. We aggregated the RFID data to a symmetric adjacency matrix $x$ where an entry $x_{ij}$ represents the number of minutes \textit{i} and \textit{j} interacted during the course of the data collection.

\paragraph{Data processing strategies}
For each of the three data processing strategies that we apply to the raw RFID data a cutoff point had to be chosen. For the first strategy, the deletion of short interactions, we chose the cutoff of 20 seconds, because none of the cutoffs tested in Study 1 improved the fit and using the cutoff of 20 seconds has been frequently applied \citep[e.g.,][]{Cattuto2010}. For the interpolation strategy, we chose to interpolate interactions that were no longer than 75 seconds apart (which produced the most sensitive result in Study 1). For the triadic closure strategy we chose to iterate once, as this produced more accurate results in Study 1 than iterating twice or more.

\paragraph{Questionnaire}
On the Sunday evening of that weekend, every participant received an email with an invitation to participate in the online survey. Among other things, we asked them ''with whom did you have pleasant interactions on the 'welcome-weekend'?''\footnote{We also asked participants with whom they had conflictive interactions. This measure was excluded from the analysis because only 9 such interactions were reported.}. 
Below this item there were 20 name generators displayed (i.e., text boxes where participants were asked to enter the names of the individuals). An auto-complete function suggested the full names of other participants when starting to type in this text field. The nominations on that item were used to construct a binary adjacency matrix $y$ where an entry $y_{ij}$ is one when an individual \textit{j} was nominated in the questionnaire by individual \textit{i}.

\subsection{Results Study 2}

\begin{figure}[ht]
    \includegraphics[scale=.7]{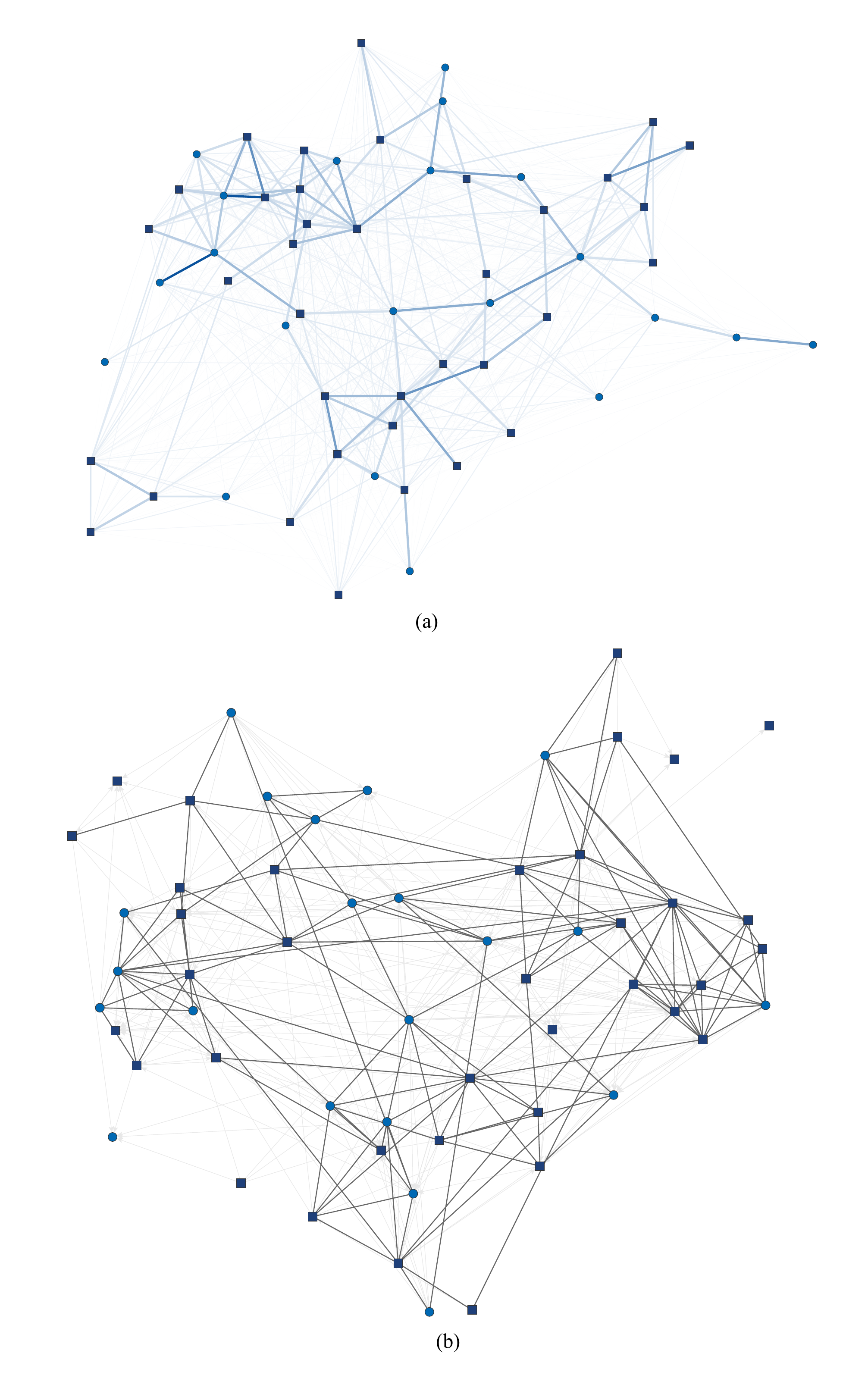}
    \caption[networks]
      {(a) weighted duration network of RFID badge signals. edge thickness = aggregated interaction duration, dark blue squares = male, light blue circles = female. (b) social interaction self-report network. dark gray ties = reciprocated ties, light gray directed = asymmetric ties, dark blue squares = male, light blue circles = female. Plotted with a backbone-layout \citep{Nocaj2015} in \textit{visone} \citep{visone}.}
    \label{networks}
\end{figure}

In total, 82.747 interactions were recorded with the RFID badges. 
157 dyads (9.84\%) did not interact at all during the course of the weekend. Figure~\ref{networks}a shows the weighted network of social interactions collected with the RFID badges.

The RFID dataset was then processed with each of the three processing functions (deletion of interactions shorter than 20 seconds, interpolated interactions that are less than 75 seconds apart, and one triadic closure iteration). Table~\ref{desc_s2} summarizes descriptive properties of these four datasets that were analyzed in Study 2. As expected, the deletion of interactions shorter than 20 seconds resulted in fewer but on average longer interactions. In a similar vein, interpolating interactions lead to longer interactions and more time spent in interactions on an aggregated level. Interestingly, the triadic closure procedure led to the creation of many interactions that were short (reflected in the low interaction duration mean in Table~\ref{desc_s2}). 

\begin{landscape}
\begin{table}[ht]
\caption{Descriptive statistics of the four RFID datasets used in Study 2\\}
\label{desc_s2}
\begin{minipage}{\textwidth}
\begin{tabular}{rrrrrrrrr}
  \hline
& \multicolumn{8}{c}{Dataset} \\\cline{2-9}
 & \multicolumn{2}{c}{unprocessed} & \multicolumn{2}{c}{< 20 duration deleted}& \multicolumn{2}{c}{interpolated w/ 75 sec}& \multicolumn{2}{c}{one triadic closure}\\ \cmidrule(r){2-3} \cmidrule(r){4-5} \cmidrule(r){6-7} \cmidrule(r){8-9} 
    
& M & \textit{SD} & M & \textit{SD} &M & \textit{SD} &M & \textit{SD}   \\ 
  \hline
  
   Interaction Duration [in sec] & 23.56 & 46.30 & 58.78 & 83.06 & 101.42 & 265.20 & 27.43 & 130.00 \\ 
  Aggregated Dyadic Duration [in min]& 15.52 & 40.30 & 9.83 & 32.60 & 22.78 & 52.89 & 36.04 & 110.41 \\ 
  Individual Total Duration [in min]& 868.87 & 442.14 & 550.39 & 357.51 & 1275.85 & 579.89 & 2018.38 & 1520.13 \\ 
   \hline
\end{tabular}
\end{minipage}
\newline
\begin{tablenotes}
      \item \newline \textit{Note.} $N_{\text{unprocessed}} = 82.747$, $N_{\text{< 20 duration deleted}} = 20.677$, $N_{\text{interpolated}} = 25.318$, $N_{\text{triadic closure}} = 158.501$, M = Mean, \textit{SD} = Standard Deviation. Interaction Duration = the  duration of an RFID interaction, Aggregated Dyadic Duration = the length of an interaction for each dyad aggregated over the whole data collection period. Individual Total Duration = the time spent in social interactions per individual aggregated over the whole data collection period.
    \end{tablenotes}
\end{table}
\end{landscape}

The self-reported interaction network consisted of 490 ties, where on average individuals nominated 9.61 (\textit{SD} = 4.03) others. 57.14\% of the ties were reciprocated. The self-report interaction network is shown in Figure~\ref{networks}b.

\begin{figure}[ht]
    \includegraphics[scale=.65]{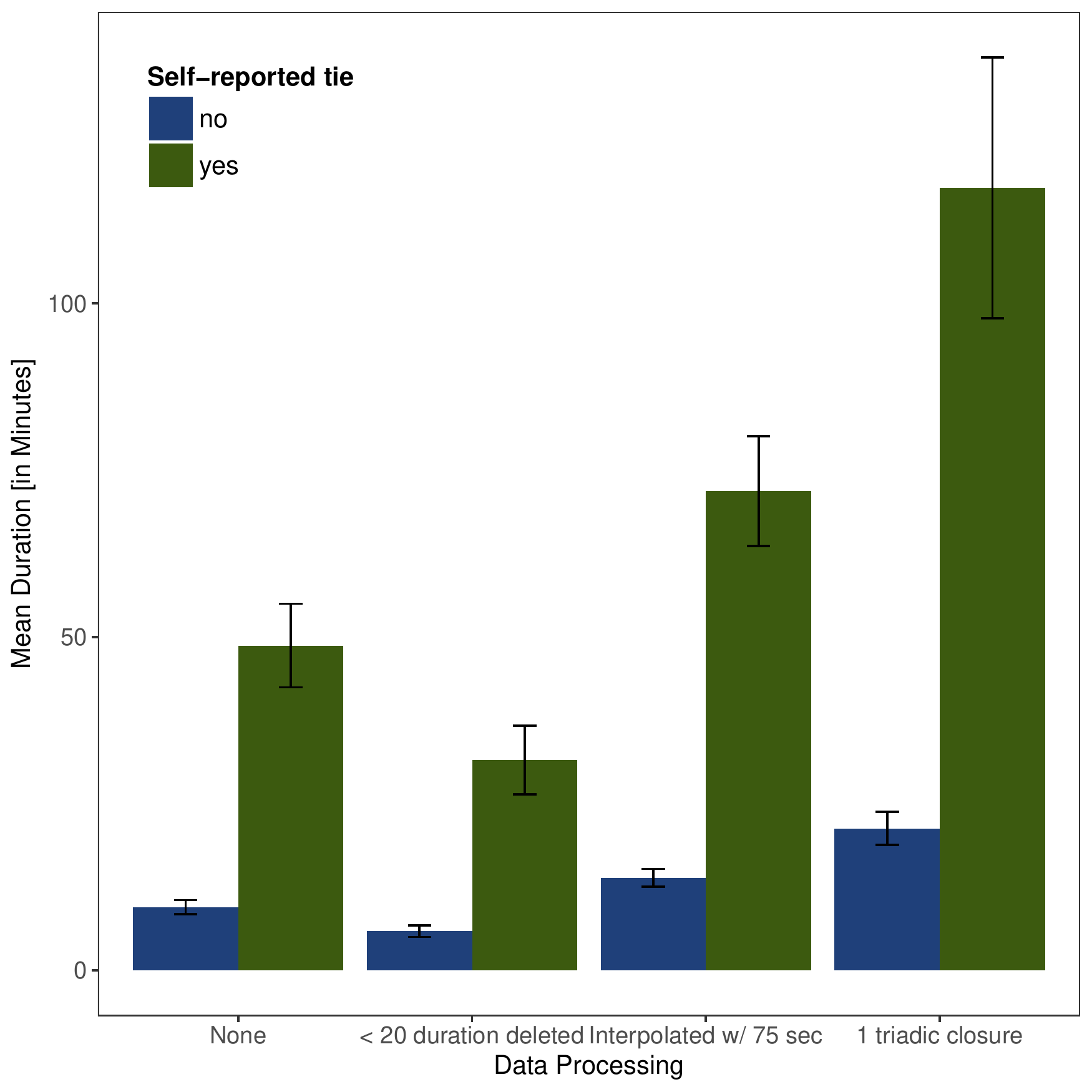}
    \caption[questionnaire + 95\% CI]
      {Means and 95\% confidence intervals of RFID interaction durations separately reported by the presence of a self-reported tie and by the applied data processing strategy.}
    \label{questionnaire}
\end{figure}

 
Figure~\ref{questionnaire} shows the mean and 95\% confidence intervals of the dyadic RFID interaction duration by the presence of a self-reported interaction and the applied data processing strategy. 
 The duration of an interaction was significantly larger if an interaction was also self-reported, t(520) = -12.10, \textit{p} < .001, \textit{d} = 1.04. This was also the case in all processed datasets (<20 seconds interactions deleted; t(520) = -9.66, \textit{p} < .001, \textit{d} = 0.82; interpolated, t(520) = -13.69, \textit{p} < .001, \textit{d} = 1.19; one triadic closure, t(520) = -13.02, \textit{p} < .001, \textit{d} = 1.09).
 
Additionally, we ran logistic regression models predicting self-reported nominations by interaction duration. The dependent variable in this analysis consisted of the vector of binary indicators for each dyad, thus $N_{dyads} = N(N-1) = 3705$, indicating whether or not a social interaction was reported in the adjacency matrix. The independent variable consisted of the duration in minutes of social interactions measured with the RFID badges over the period of the data collection. Table~\ref{log} shows the results of these models for the raw dataset and for the datasets that have been processed with one of the three data processing strategies. In all datasets the duration of the interaction predicted self-reports of interactions. A good indicator of the fit between the RFID data and the self-reports of social interaction is McFadden's pseudo $R^2$ (i.e., how much variance in the dependent variable each model explains). The highest $R^2$ is achieved for the model in which the RFID data have been interpolated with a cutoff of 75 seconds. A likelihood ratio test comparing the models of the interpolated dataset to all other models indicates that the model with the interpolated data outperforms all other datasets (comparison with unprocessed data: $\chi^2(2,3192) = 77.3, p < .001$, data where <20 sec interactions are deleted: $\chi^2(2,3192) = 184.0, p < .001$, dataset with one iteration of triadic closure: $\chi^2(2,3192) = 75.9, p < .001$).

 \begin{landscape}
\begin{table}[ht]
\caption{Logistic Regression Models on the self-reports of social interactions}
\begin{tabular}{rrlrrlrrlrrlr}
\hhline{=============} 
 &  \multicolumn{12}{c}{Dataset} \\\cline{2-13}
& \multicolumn{3}{c}{unprocessed} & \multicolumn{3}{c}{< 20 duration deleted} & \multicolumn{3}{c}{interpolated w/75 sec} & \multicolumn{3}{c}{one triadic closure}\\ \cmidrule(r){2-4} \cmidrule(r){5-7} 
\cmidrule(r){8-10}\cmidrule(r){11-13}  
 & Est. & & S.E. & Est. & & S.E.& Est. & & S.E.& Est. & & S.E.  \\ 
  \hline
Intercept & -2.16 &*** & 0.06  & -2.00&*** & 0.06  & -2.28&*** & 0.06  & -2.18&*** & 0.06  \\ 
  RFID Duration & 0.02&*** & 0.00  & 0.02&*** & 0.00  & 0.02&*** & 0.00  & 0.01 &*** &  0.00 \\ 
  McFadden's $R^2$ & .12 &  &  & .08 &  &  & .14 &  &  & .12 &   \\ 
  Log Likelihood & -1209.43 & &   & -1262.76 & &   & -1170.77 &  &  & -1208.71 &   \\ 
\hhline{=============} 
\newline
\end{tabular}
\newline
\begin{tablenotes}

      \item \textit{Note.} N = 3192, * \textit{p} $<$ .05, ** \textit{p} $<$ .01, *** \textit{p} $<$ .001
    \end{tablenotes}
    \label{log}
\end{table}
 \end{landscape}

Moreover, we tested the robustness of these findings against alternative definitions of the criterion variable (i.e., the self-reports), as this measure might suffer from various biases associated with self-report measures \citep[e.g., recency effects;][]{Greene1986}. In these additional analyses, we considered interaction to be reported if a) at least one of the two individuals reported the interaction (weak symmetrization) or b) if both individuals reported the social interactions (strong symmetrization). Appendix~\ref{app:sym} shows the results of these analyses, indicating that also for alternative definitions of the criterion variable there is a strong overlap with the RFID data and that the interpolation strategy increases the model fit. 

\subsection{Discussion Study 2}

In Study 2 we have shown that face-to-face interaction recorded with the RFID badges correspond highly with self-reports of social interactions. Moreover, we have shown how the overlap between the data recorded by RFID badges and self-reports changes based on the application of three different data processing strategies: deleting interactions that are shorter than 20 seconds, interpolating interactions that are no longer than 75 seconds apart (which produced the most sensitive result in Study 1), and one iteration of triadic closure.  When applying the interpolation and triadic closure data processing strategies we observed an increase of the fit between the RFID data and the self-reports of social interactions. The deletion of interactions that are shorter than 20 seconds lead to a decrease in the overlap. 
Compared to the raw data and all other data processing strategies, the interpolated dataset indicated the largest effect size and explained variance for the fit between the RFID data and self-reports of social interactions. Hence, RFID badges show a good criterion validity, even more so, when processed with a interpolation time of 75 seconds.
A unique feature of this analysis is that we validated RFID interactions against subjectively important interactions -- hence mixing behavioral and self-report measures -- in a sample with a compliance rate of 100\% for the RFID data collection part.

The results of Study 2 are in line previous studies on the overlap between RFID data and self-report data on social interactions \citep{Thiele2014,Atzmueller2018}. Together with our results, these findings indicate that there is a large but not perfect overlap between RFID measures and self-reports, indicating that each method still measure something unique that the other does not capture. 
Our study, however, goes one step further and compares the overlap of these methods for all three strategies that aim at enhancing the validity.

Study 2 has some limitations. First, we asked the participants with whom they had \textit{pleasant} interactions. Individuals might also have had negative or neutral interactions which might have biased our analysis. However, due to the unstructured nature of this social event, we believe that individuals tended to engage and stay in social interactions that they perceived as pleasant. We thus think that this bias is very small. Second, individuals tend to be biased in retrospective reports \citep{Bernard1984}. More specifically, recall, recency, and an alcohol-related memory bias might have affected the self-report data in a way that favored interactions that were emotionally arousing \citep{Mather2011}, happening later on the weekend \citep[recency bias;][]{Greene1986}, or those that happened in the absence of alcohol intoxication \citep{Sullivan2010a}.

\section{General Discussion and Conclusion}

In two studies we assessed the construct and criterion validity of RFID badges to measure social interactions. We conclude that RFID badges are in part a valid measure of social interactions. On the one hand, the construct validity (as measured by sensitivity and specificity) is not very high but can be enhanced to an acceptable level when merging interactions that are no longer than 75 seconds apart. 
We considered it promising that the specificity of the measure tends to be very high (96\%) while the sensitivity is rather conservative (about 50\%). If the RFID sensors capture a signal this is likely to be a true face-to-face interaction. 
We further found that the face-to-face interactions recorded with the RFID badges are highly associated with self-reports of social interactions (criterion validity). 

This study aimed to fill a research gap by providing validity tests of RFID badges to measure social interactions. RFID badges are a technology increasingly used in behavioral research studies due to their easy applicability, relatively low costs, and moderate privacy intrusion as compared to other technologies. 

Since RFID badges can be applied in settings where measuring social interactions is otherwise difficult, future studies could, for instance, investigate how social interactions lead to meaningful social relations (e.g., friendship ties), how individuals informally form social groups, or how social behaviour interplays with individual’s cognitions and feelings. Particularly in combination with other measurement methods (e.g., surveys), the application of RFID badges can offer new insights into individuals' experiences and behaviours.
Technically, the RFID badges could be expanded to work without readers (i.e., the badges would store the data), thus allowing more flexible and larger spatial settings. Also, the badges can easily be combined with other measurement devices \citep[e.g., a microphone;][]{Lederman2017} to measure additional aspects of social interactions. Additional measurement devices, however, increase the privacy intrusiveness and thus reduce the ease of application.

 
This study contributes to our knowledge of what RFID badges measure and under which circumstances their application is warranted. We believe that RFID badges and other types of sociometric badges hold great potential to improve our understanding of how individuals engage in social interactions \citep{Pentland2008}. In our view, RFID badges are a promising tool to evaluate ''actual'' individual behavior in the psychological sciences \citep[as recently called for by][]{Baumeister2007} 
and can help to gain new insights into the crucial effects of human face-to-face interactions.

\newpage
\renewcommand{\refname}{References}

\bibliography{extracted}

\newpage
\section*{Appendix}
\renewcommand{\thesubsection}{\Alph{subsection}}

\subsection{Study 1: Classification table for the interpolation strategy of with a 75 seconds cutoff}
\begin{table}[ht]
  \caption{Classification table (in seconds) between the RFID data and the video-coded data for the interpolation strategy of with a 75 seconds cutoff}
  \label{stats75}
  \begin{minipage}{\textwidth}
\begin{tabular}{
L{2cm}
L{2cm}
L{2cm}
L{2cm}
}
\hhline{====} 
\\
& & \multicolumn{2}{c}{RFID} \\
& & positive & negative \\
 \hline \\
Video & positive & 33,446 & 17,554 \\
& 	negative & 10,638 & 191,473 \\
\hline

\hhline{====}
\end{tabular}
 \end{minipage}
\end{table}

\subsection{Study 1: Alternative agreement measure}\label{app:S1_robust}
\begin{figure}[ht]
    \includegraphics[scale=0.65]{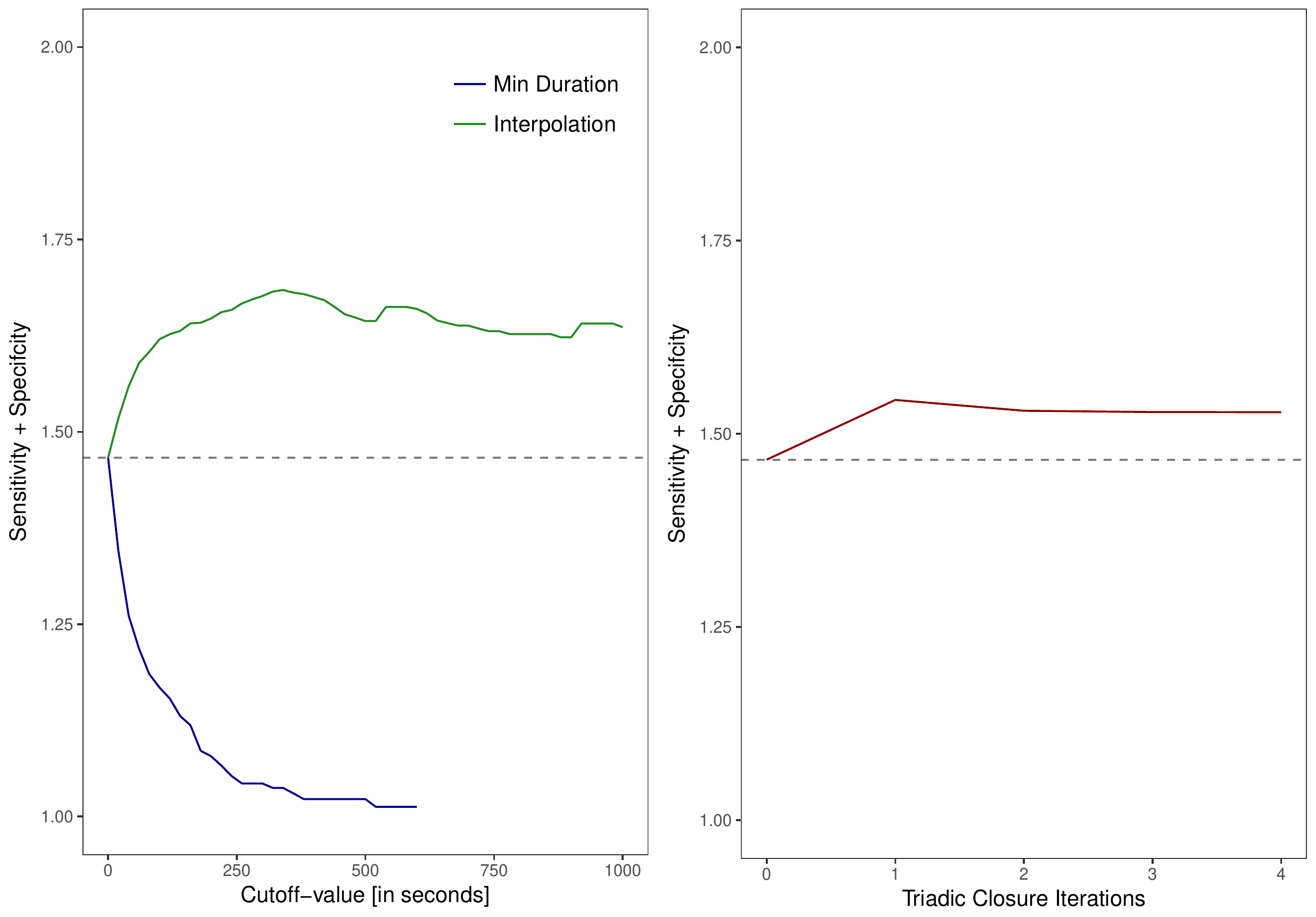}
    \caption[questionnaire + 95\% CI]
      {Sum of the sensitivity and specificity of the RFID data compared to the video coded data after the two strategies (minimal duration and interpolation) have been applied with different cutoff-values (left plot) and after several rounds of triadic closure iterations (right plot). The horizontal gray line indicates the sum of the sensitivity and specificity of the unprocessed data.}
    \label{supp2}
\end{figure}

In Study 1 we optimized the fit between the RFID data and the video data based in the accuracy agreement measure. One could also choose other measures to assess this fit. In the analyses reported here, we considered another measure of fit - the sum of the sensitivity and specificity. Figure~\ref{supp2} shows the sum of the sensitivity and specificity of various cutoff-points for the three data processing strategies.

The highest sum of sensitivity and specificity was obtained with the interpolation strategy and a cutoff of 340 seconds. In our view, interpolating over 5 minutes of gaps between two RFID signals is too much and would drastically decrease the fine-grained quality of the data. Also, please note that the sum of sensitivity and specificity is a relative measure (as it indicates the relative proportion of correctly identified interactions or non-interactions, respectively) and does not treat each specified second equally. For this reason, we focused our analysis in the main text on optimizing the accuracy as a non-relative measure.

\subsection{Study 2: Alternative definitions of the criterion variable}
\label{app:sym}

\begin{figure}[ht]
    \includegraphics[scale=.65]{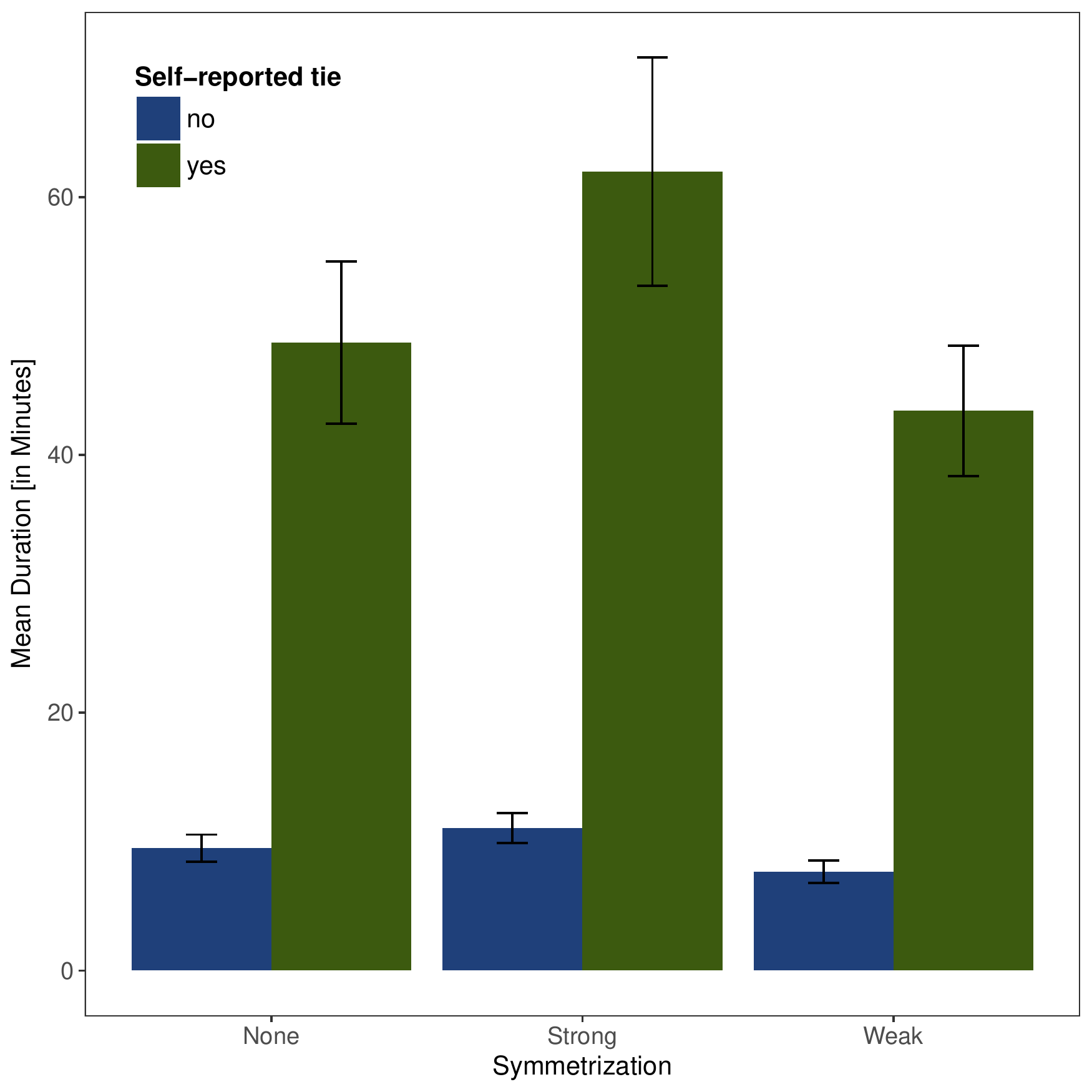}
    \caption[questionnaire + 95\% CI]{}
      {Means and 95\% confidence intervals of RFID interaction durations separately reported for the non-symmetrized, strongly symmetrized, and weakly symmetrized self-reports.}
    \label{mean_sym}
\end{figure}

In additional analyses reported here, we consider the asymmetry and mutuality of self-reports of social interactions. On the one hand, it might be a more reliable indicator of a social interaction if both individuals of a dyad reported it as such. Hence, we would only consider mutual reports (i.e., Person A reports an interaction with Person B and Person B reports an interaction with Person A). On the other hand, one could argue that asymmetric reports (only one of the two reported the interaction) might come about by recall problems of self report and other measurement biases, e.g., because someone does not remember the name of the other when reporting social interactions. For these reasons, we have conducted further analyses in which we consider an interaction to be reported if a) at least one of the two individuals reported the interaction (weak symmetrization) or b) if both individuals reported the social interaction (strong symmetrization). 
Figure~\ref{mean_sym} shows the average RFID durations by the presence of a self-reported interaction and by the symmetrization processing. In all cases, the mean of the RFID durations was higher for those interactions that were also self-reported but effect sizes did not differ much between the different symmetrization strategies (no symmetrization: t(520) = -12.08, \textit{p} < .001, \textit{d} = 1.04, weak symmetrization: t(520) = -15.53, \textit{p} < .001, \textit{d} = 1.08, strong symmetrization: t(520) = -8.96, \textit{p} < .001, \textit{d} = 1.05).

Table~\ref{rob1} shows the results of three logistic regression analyses in which the dependent variables were either the unprocessed self-reports or the weakly or strongly symmetrized data. The independent variable was always the duration of the interactions measured with the RFID badges.

\begin{table}[ht]
\caption{Logistic regression analyses on three different symmetrization strategies of self-reported interactions as dependent variables}
\label{rob1}
\begin{tabular}{rrlrrlrrlrrlr}
\hhline{==========} 
 &  \multicolumn{9}{c}{Symmetrisation} \\\cline{2-10}
& \multicolumn{3}{c}{None} & \multicolumn{3}{c}{Weak } & \multicolumn{3}{c}{Strong}\\
\cmidrule(r){2-4} \cmidrule(r){5-7} 
\cmidrule(r){8-10} 
 & Est. & & S.E. & Est. & & S.E.& Est. & & S.E. \\ 
  \hline
Intercept & -2.16 & *** & 0.06 & -1.81 & *** & 0.06 & -2.81 & *** & 0.08 \\ 
  RFID Duration & 0.02 & *** & 0.00 & 0.03 & *** & 0.00 & 0.02& ***  & 0.00 \\ 
  McFadden's pseudo $R^2$ &0.12 &  && 0.14 &  & &0.13 &  \\ 
  Log Likelihood & -1209.43 &  && -1445.02 &  && -828.88 &  \\ 
   \hline
\end{tabular}
\newline \\

\begin{tablenotes}
      \item \textit{Note.} N = 3.192, * \textit{p} $<$ .05, ** \textit{p} $<$ .01, *** \textit{p} $<$ .001
    \end{tablenotes}

\end{table}
   
When comparing the log-likelihoods of the three models in Table~\ref{rob1}, we see that the model with the strong symmetrization outperforms the other models (with no symmetrization: $\chi^2(2,3192) = 761.1, p < .001$, with weak symmetrization:$\chi^2(2,3192) = 1232.3, p < .001$). The model with the weakly symmetrized dependent variable has a lower log-likelihood than the one with the non-symmetrized dependent variable, $\chi^2(2,3192) = 471.2, p < .001$. This suggests that, indeed, there is some measurement error in the self-reported that can be reduced by only considering mutual nominations.
    

\subsection{Study 2: Rank based comparison}

\begin{figure}[ht]
    \includegraphics[scale=.65]{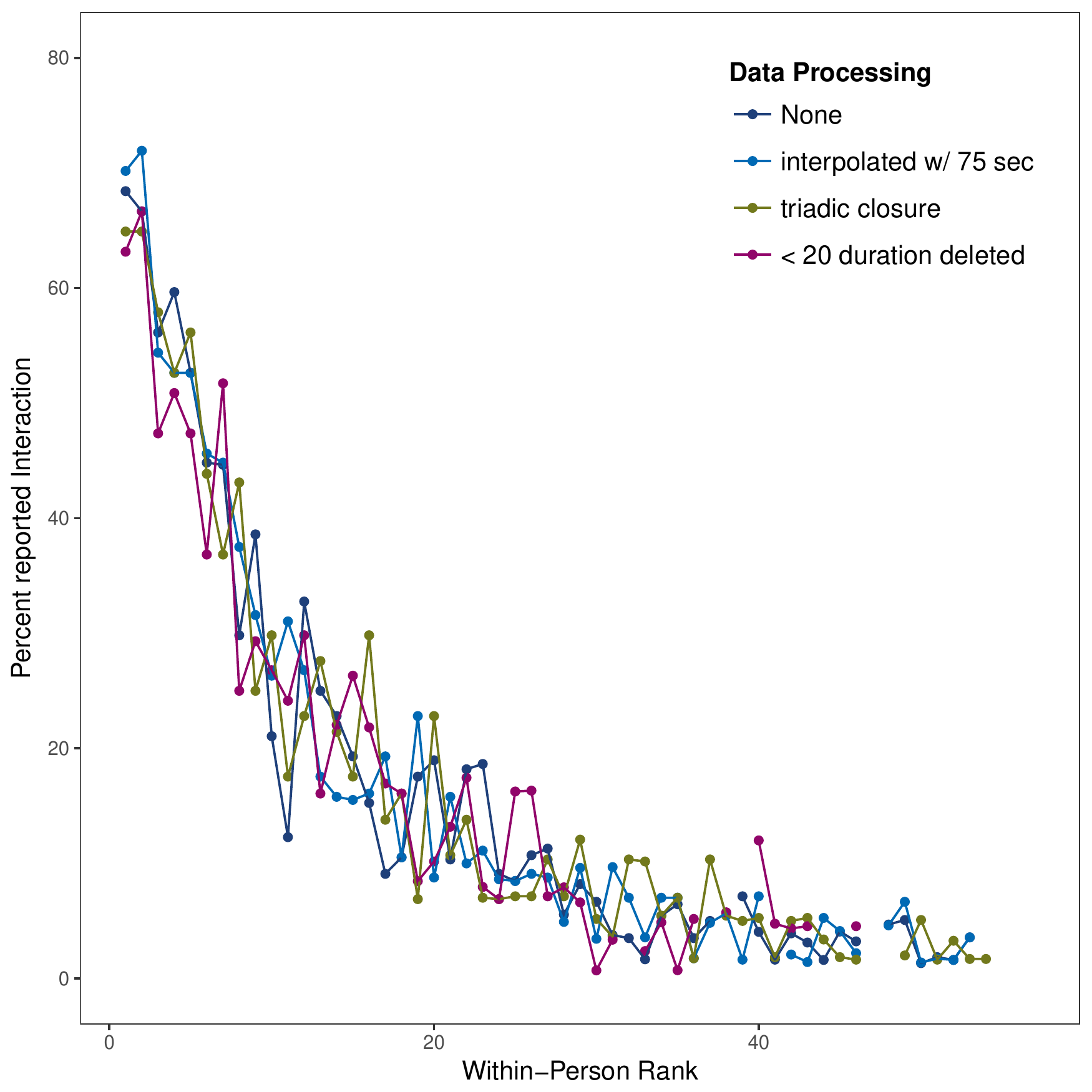}
    \caption[hits]
      {The percentage of reported interaction by the within-person rank of the interaction duration and by the type of data processing. Some ranks are not computed due to rank-ties with other ranks.}
    \label{hits}
\end{figure}

To account for the inter-individual variability in interaction durations over the data collection, we computed the within-person rank order of RFID interactions and compared this to the self-report of interactions. In other words, we ranked -- for each individual -- the interactions with all other individuals in decreasing order (i.e., the longest duration was rank one). Then, we computed the percentage of interactions of a given rank that were reported as social interactions. Figure~\ref{hits} shows these percentages by rank and data processing strategy. None of the three data processing strategies show a clear improvement compared to the raw dataset. A clear improvement would be visible by higher percentages for the lower ranks and lower percentages for the higher ranks.

\section{Acknowledgments}
We thank the participants of both studies for their support and trust, the \textit{Swiss StudentLife} team at ETH Zurich, the Social Networks group at the ETH Zurich, Stefan Wehrli and the DeSciL lab at ETH Zurich, Julia von Fellenberg, Kieran Mepham, Ulrik Brandes, and a particular student organization.

\end{document}